\documentclass[mathleft
]{an}
\usepackage[T1]{fontenc}
\usepackage{ae,aecompl}
\usepackage{graphicx}
\usepackage{times}
\usepackage{natbib}
\newcommand{\hi}{{H\sc{i}~}}

\begin{document}

\Pagespan{1}{}
\Yearpublication{2011}%
\Yearsubmission{2010}%
\Month{11}%
\Volume{999}%
\Issue{88}%

\title{The Effelsberg Bonn \hi Survey (EBHIS)}

\author{J\"urgen Kerp\inst{1}\fnmsep\thanks{Corresponding author:\email{jkerp@astro.uni-bonn.de}\newline}, Benjamin Winkel\inst{1}\fnmsep\thanks{current affiliation: Max-Planck-Institut f\"ur Radioastronomie, Auf dem H\"ugel 69, D-53121 Bonn}, Nadya Ben Bekhti\inst{1}, Lars Fl\"oer\inst{1} and Peter M.W. Kalberla \inst{1}}
\institute{Argelander-Institut f\"ur Astronomie, Auf dem H\"ugel 71, 53121 Bonn, Germany}

\titlerunning{Effelsberg-Bonn \hi Survey}
\authorrunning{Kerp et al.}

\received{5 Nov 2010}
\accepted{23 Feb 2011}
\publonline{later}


\abstract{The Effelsberg-Bonn \hi survey (EBHIS) comprises an all-sky survey north of Dec $ = -5\degr$ of the Milky Way and the local volume out to a red--shift of $z \simeq 0.07$. Using state of the art Field Programmable Gate Array (FPGA) spectrometers it is feasible to cover the 100\,MHz bandwidth with 16.384 spectral channels. High speed storage of \hi spectra allows us to minimize the degradation by Radio Frequency Interference (RFI) signals.  Regular EBHIS survey observations started during the winter season 2008/2009 after extensive system evaluation and verification tests. Until today, we surveyed about 8000 square degrees, focusing during the first all--sky coverage of the Sloan-Digital Sky Survey (SDSS) area and the northern extension of the Magellanic stream. The first whole sky coverage will be finished in summer 2011. Already this first coverage will reach the same sensitivity level as the Parkes Milky Way (GASS) and extragalactic surveys (HIPASS).
EBHIS data will be calibrated, stray--radiation corrected and freely accessible for the scientific community via a web-interface.
In this paper we demonstrate the scientific data quality and explore the expected harvest of this new all--sky survey.}

\maketitle

\section{Introduction}
\hi is the simplest and most abundant atom in space. Using todays radio astronomical technology, \hi is easy to detect. The sensitivity of the world's largest radio telescopes allows us to measure the neutral species down to volume densities where a major gas fraction is already ionized \citep{wolfire1995a}. \hi is of scientific key-importance for our understanding of galaxy formation, evolution and merging, because even at large distances from the stellar body we can study in great detail the density, temperature and velocity structure of the neutral gas \citep{kalberlakerp2009}.

Galactic \hi covers the whole sky! The Lockman area \citep{lo86} and the Chandra deep--field south area \citep{braha2005} denote minimum column density regions where the gaseous medium reaches local minimum column densities, these are the windows for high--energy astrophysics. Here the attenuation of the highly red--shifted objects is low and smoothly distributed. Despite the fact that neutral hydrogen itself has not the largest photo--electric cross section for i.e. X-rays, it is the quantitative tracer for the spacial distribution of heavier species. \hi full--sky single dish surveys are accordingly of high importance to quantify the amount of matter distributed along the line of sight across the whole sky. Towards high galactic latitudes the \hi column density value itself is a measure for this quantity, towards the inner Galactic plane, the radial velocity information allows to disentangle the spatial composition of different portions of the Galactic disk. Also the distant outskirts of the Milky Way can be explored by means of the \hi radiation. Here sophisticated models are needed to disentangle the spacial distribution of the gaseous layers. Combining all this information it is feasible to disclose the distribution of gravitational matter far beyond the stellar disk \citep{kalb2007}.

EBHIS is the first all--sky survey which aims to perform a blind survey of the Milky Way \hi distribution and the local volume in parallel. Present day FPGA spectrometer \citep{stanko2005, klein2009} allows us to resolve the Milky Way cold neutral medium (CNM) spectral lines and to measure the \hi radiation of a $M({\rm HI}) \geq 10^9\,{\rm M_\odot}$ galaxy at a red--shift of 0.07 within a single spectrum. Making use of this ability we optimized EBHIS to cover the full northern sky at an unique signal--to--noise ratio (SNR). Here, full angular sampling and the high sensitivity of the Effelsberg 100-m dish are of equal importance for the data quality. Towards the Sloan--Digital--Sky Survey (SDSS; \citet{SLOAN2000}) area we optimized EBHIS to provide a complete census of the \hi emission of $M(HI) = 3\cdot 10^7\,{\rm M_\odot}$ galaxies at the distance of the Virgo cluster. While on cosmological scales the Universe is considered to be isotropic and homogeneous, we know that the local volume towards the northern hemisphere offers a unique laboratory towards the closest large galaxy clusters, our large and massive local group spiral galaxies and a huge variety of high--velocity cloud (HVC) complexes which show--up with unique signatures of ongoing interaction with the Milky Way Galaxy.
  
\subsection{Milky Way Surveys}
Today the main resource for full sky \hi studies of the Milky Way is the Leiden/Argentine/Bonn (LAB) survey \citep{kalb2005}. This survey comprises the data of two major surveys, the Leiden/Dwingeloo survey \citep{hartmannburton1998} and the Instituto Argentina de Radioastronomia (IAR) survey \citep[][]{arnal2000,bajaja2005}. Both surveys are performed with 25-m or 30-m single dish radio telescopes, respectively. To survey the whole accessible sky in a reasonable amount of time with a single feed receiver the angular sampling of the sky was chosen to be on a  ``beam-by-beam'' grid. Following the Nyquist sampling theorem the effective angular resolution is approximately $1\degr$.

The LAB survey represents the most coherent \hi data\-base of the whole Galactic sky today, superior because of its sensitivity (70\,mK), completeness and correction for stray--radiation \citep{kalb1980, kalb2005}. It replaced the Bell-Labs survey (Stark et al. 1992) as the standard resource of information on the Milky Way neutral gas distribution.
The velocity resolution of the LAB-survey ($\Delta v = 1.25\,{\rm km\,s^{-1}}$) is sufficient to resolve narrow lines associated with the \linebreak CNM (FWHM $< 4\,{\rm km\,s^{-1}}$, corresponding to \linebreak $T_{\rm gas} \simeq 300$\,K, \citet{haud2007}). The velocity coverage of LAB ($-450 < v_{\rm LSR} < 400\,{\rm km\, s^{-1}}$) allows to detect almost all emission belonging to the Milky Way galaxy, considering that the escape velocity of the Milky Way gravitational potential is about $540\,{\rm km\,s^{-1}}$. During the past two years the Parkes 21-cm multi-feed receiver system was used to perform a full sky survey south of Dec(2000) = $0\degr$. The so-called Galactic All-Sky Survey (GASS, \citet{naomi2009, kalb2010}) provides an angular resolution of 15.6 arcmin, a velocity resolution of \linebreak $0.8\,{\rm km\,s^{-1}}$ and a root--mean-square (rms) sensitivity of \linebreak about 60\,mK.

EBHIS and GASS open new opportunities for the Milky Way ISM studies, for the following reasons: 
\begin{enumerate}
\item Complete census of all HVCs: as discussed below, towards the northern hemisphere ultra--compact HVC \linebreak have only be accidentally identified. Some HVC head-tail structures \citep{bruens2000} might be wrong\-ly i\-den\-ti\-fied due to insufficient angular resolution.
\item Multiphase structure of the extra-planar gas: towards the northern sky we found evidence for the interaction of HVCs with IVCs and Milky Way gas \citep{kerp99, kappes2003}. Because of the factor seven higher angular resolution of EBHIS in comparison to LAB, we will gain statistical evidence for this ``fueling'' process of the Milky Way ISM.
\item \hi mass spectrum of clouds: Absorption line studies by \cite{nadya2010} disclose the existence of cold and dense neutral gas clumps embedded within a low volume density environment. EBHIS will allows us to establish a complete and homogenous data--base for the cloud \hi mass and size spectrum.
\item \hi shells: feedback processes between stellar evolution and the ISM are important for our understanding of the cycle of matter. EBHIS will complement towards low galactic latitudes the radio interferometric surveys \linebreak \citep{Taylor2003, naomi2001} of the Milky Way and allows us to quantify the galactic fountain model.
\item Soft X-ray absorption: EBHIS will serve as a unique data base for about a decade or more to evaluate the strength of photoelectric absorption of soft X-ray photons. Focusing towards high Galactic latitudes, soft X-ray attenuation is dominated by the warm neu\-tral me\-di\-um (WNM) \citep{kerp2003}.
\end{enumerate}

\subsection{Extragalactic Surveys}
With the advent of multi-feed receivers in radio telescopes it has become possible to map large portions of the sky on an fully-sampled angular grid with a fraction of the observing time necessary compared to a single feed receiver system. The prime example is the \hi Parkes All Sky Survey (HIPASS, \citet{Barnes2001}) which represents a major leap in extragalactic radio astronomy. Similar projects were initiated at the Jodrell Bank radio telescope (HI Jodrell All Sky Survey (HIJASS), \citet{lang2003}) and the Arecibo 300-m dish \citep{gio2005}. While HIPASS and HIJASS (which was never finished) are comparable in all relevant observational parameters, the Arecibo dish equipped with the Arecibo L-Band Feed Array (ALFA) is unique in its sensitivity and angular resolution and used for a variety of individual scientific projects to map the \hi distribution towards the sky accessible from Arecibo i.e. the Galactic (GALFA, \citet{Stani2006}) and extragalactic surveys \citep{gio2005}. Deeper surveys are in progress towards selected areas, for example with $T_{\rm int} \simeq 300\,$s per position 200 sq. deg will be observed (ALFALFA,\linebreak \citet{gio2005}), the ALFA Ul\-tra Deep Sur\-vey \linebreak (AUDS, \citet{freud2005}) will map 0.4 sq. deg with 75 hours of integration time, and a survey of the accessible part of the Zone Of Avoidance (ZOA, \citet{henning2010}) has been initiated.
Surveys are improving our knowledge of fundamental questions on
\begin{enumerate}
\item The low-mass part of the \hi mass function: the EBHIS SDSS area (10\,min. per position) will offer an \hi mass sensitivity of \linebreak $M(HI) = 3\cdot 10^7\,{\rm M_\odot}$ at the distance of the Virgo cluster. EBHIS mass sensitivity and the richness of nearby large-scale structures towards the northern sky, we expect to increase the number of $M(HI) < 10^8\,{\rm M_\odot}$ ga\-la\-xies \citep{zwaan2005} by an oder of magnitude.  
\item The local baryon budget: considering the general direction of the constellation Coma Berenices EBHIS will detect the \hi content of 274 galaxies, 37 galaxy groups and three galaxy clusters with a mass sen\-si\-ti\-vi\-ty \linebreak of $10^7\,{\rm M_\odot}$ ($v_{\rm LSR} \leq 1000\,{\rm km\,s^{-1}}$). Implying a highly statistical significant census of \hi in the local universe.
\item High mass \hi galaxies: high quality \hi data of bright galaxies due to high spectral resolution and the sensitivity.
\item Isolated \hi clouds in the intergalactic medium: towards the Taurus, Leo, Gemini and Coma Void ($D \leq 55\,{\rm Mpc}$) we expect to reach an \hi mass sensitivity of about \linebreak $M(HI) \leq 4\cdot 10^8\,{\rm M_\odot}$ -- comparable to the \hi mass of the Large Magellanic Cloud.
\item Search for galaxies close to low red--shift Ly--$\alpha$ ab\-sor\-bers: completeness of EBHIS in combination with full angular sampling allows to identify \hi counterparts of the absorption lines on a high signal-to-noise ratio.
\item The imprint of environmental conditions on galaxies: comparison between high gravitational environments \linebreak (i.e. Perseus-Pisces and Coma super-cluster) and the adjacent voids.
\end{enumerate}
The northern hemisphere allows to study unique structures, galactic as well as extragalactic. In example the close located large scale structure towards the Coma cluster consists of about 7000 galaxies, 336 groups and 82 clusters of galaxies below $v_{\rm LSR} = 10.000\,{\rm km\,s^{-1}}$. High- and Intermediate Velocity Clouds cover a huge portion of the sky towards the northern Galactic pole. Together with GASS and HIPASS, EBHIS will be a rich resource of the whole \hi sky, providing a unique data base for radio and multi-frequency astronomy in general. The Effelsberg \hi 21-cm line survey can be thus considered as a legacy project.

\subsection{Angular Resolution, Sampling and high--speed Dumping}\label{labvsebhis}
There are three key “ingredients” that are critical for a successful performance of EBHIS 1) angular resolution, 2) full sampling and 3) high-speed dumping. We will discuss those in the following.

One of the main arguments for a new single dish full--sky survey is about an order of magnitude higher angular resolution in comparison to previous northern sky surveys like LAB. It is important to note that angular resolution is not only needed to increase the detailed structure of detected features –- perhaps more importantly -- is the aim to detect clouds and filaments that remain undetectable in lower--resolution surveys. To illustrate this, we compare in Fig.\,\ref{labvsebhis} observations towards the same portion of the sky (${\rm RA(2000)} = 19^{\rm h} 53^{\rm m}$ and ${\rm Dec(2000)} = 00\degr 30\arcmin$). The rms--noise of both surveys is comparable. Displayed in \newline Fig.\,\ref{labvsebhis} is in the top panel the ($-400 \leq v_{\rm lSR}[{\rm km\,s^{-1}}] \leq 400$) LAB spectrum and by an offset of 50\,K the EBHIS spectrum covering the same velocity range. The spectrum is extracted towards an ultra-compact high-velocity cloud (UCHVC) belonging to HVC complex GCN. In Fig.\,\ref{labvsebhis} bottom panel we display a superposition of the UCHVC line spectrum. \linebreak Shown for the LAB survey is the nearest located \hi spectrum towards the EBHIS position. The LAB spectrum does not show-up with any signal of this cloud, while EBHIS receives a $T_{\rm B} \simeq 2$\,K line . 


The striking difference between both \hi spectra is on one hand because of the difference in size of both survey radio telescopes, the beam--filling factor.
The beam-filling factor is determined by the ratio of the angular extent of the cloud and the angular resolution of the dish $f_{\rm B} = \frac{{R_{\rm cloud}}^2}{{R_{\rm tele}}^2}$. The obvious difference in the significance of the UCHVC is hence determined by the square of the ratio of the angular resolving power of both telescopes. In the extreme case --  also Effelsberg do not resolve the UCHVC -- it is a factor of 16 ($R_{\rm eff} = 9\arcmin$ and $R_{\rm LAB} = 36\arcmin$)! In fact, LAB should detect a 125\,mK signal from the UCHVC, but we are faced with a non--detection. 

We have to consider here, that on the other hand the angular sampling is of high importance. The Dwingeloo telescope did not observe the UCHVC directly, meaning the main--beam did not point exactly towards the cloud location. The main--beam was offset from the column density maximum because of the beam--by--beam sampling. Considering the fact that the LAB survey data offers in fact an angular resolution of $1\degr$. This means within an angular area of $1\degr \times 1\degr$, 278 square arcmin are not observed! At least 8\% of the whole northern area is today not observed at all. An UCHVC of $2\arcmin$ diameter covers only 1.1\% of the unobserved area. Accordingly, the detection probability of an UCHVC is at least a factor of 20 lower for LAB than for EBHIS. In conclusion, beam-by-beam sampling represents the most efficient way of carrying out an all-sky survey with a single receiver in a reasonable period of time. However, based on the \hi surveys today available we might have simply overlooked an interesting and perhaps a significant fraction of baryonic matter within the Milky Way. In fact, the current detections of UCHVCs have only been accidentally towards the northern hemisphere!

\subsection{Single Dish vs. Interferometric Surveys}
Up to now several surveys of the Galactic \hi emission have been obtained with radio interferometers (for a review see \cite{naomi2004}). A radio interferometer acts like a filter in the angular domain (spatial-frequency). The signals of the individual radio telescopes are correlated by a central processor to improve the angular resolution corresponding to the maximum linear separation of two antennas of the array. Because of the correlation process, however, the interferometric data do not completely recover information on the large-scale structure of the interstellar medium at angular scales of degrees or larger. The smallest separation between the radio dishes determines the lowest angular resolution power, at least the diameter of the individual dishes i. e. for the Australia Telescope Compact Array (ATCA) the largest detectable structure is about 27 arcmin.
Astrophysically this is of high importance for extended objects in the Milky Way or the local universe. As a prime example we refer to the observation of an HVC shown in Fig. 1 of \citet{bruens2001}. It shows a superposition of a Westerbork radio synthesis telescope (WSRT) observations superposed by white contour lines on a gray-scale map obtained with the 100-m Effelsberg telescope. Because of the correlation procedure applied to the WSRT data only the cold gas core with high column densities was detected. The enveloping WNM, however, is ``filtered out'', undetectable by any radio interferometer. Scientifically, however, the WNM contains the important information in this particular case. The HVC is most likely stripped-off by ram-pressure interaction with the ambient medium and heated up from the cold core to its rims. In this particular case the Effelsberg observation discloses the fact, that this HVC interacts with the low volume density medium in the transition region between the Galactic Halo and the intergalactic medium of the local group of galaxies. The cold core, resolved by the WSRT, is also detected by the single dish data. But in addition to the radio interferometer, the Effelsberg data allows us to disentangle the contribution of {\em all} individual gas phases across the whole area of interest.
This implies that the gas of the HVC is heated up by the interaction process. This allows us to determine the heat capacity and thermal conductivity of the gas within the cloud, which is impossible to achieve with the radio interferometer data alone! This example stresses the importance of single dish observations of extended structures, not restricted to the gaseous distribution of the Milky Way galaxy only.

\section{Instrumental Setup}
In this section we compile the basic information on the \newline EBHIS instrumental setup. Details on the system characteristic and performance can be found by \citet{winkel2010}.
\subsection{The Receiver}
The  21-cm seven beam receiver was primarily built for \linebreak beam-\-park experiments to measure distribution and size \linebreak spectrum of space debris. The tracking and imaging radar system (TIRA) antenna of Fraunhofer-Institut for high frequency physics and radar technique in \linebreak Wacht\-berg--Wert\-hofen near Bonn is used as transmitter, the Effelsberg 100-m telescope as receiver, working as bistatic radar arrangement. Particles down to 9\,mm size can be detected at altitudes between 800 to 1000\,km.

The 21-cm multi-feed receiver is a cooled single conversion heterodyne system offering 14 separate receiving channels. The electromagnetic waves are focused by the antenna and couple via the feed horns into waveguides placed in a cryogenic dewar. The whole receiver fits in into a single receiver box and is situated in the primary focus of the telescope. The central feed is sensitive for circular polarization while the off-set feeds receive linear polarized signals.

\subsection{The Spectrometer}
EBHIS was designed to survey in parallel the Milky Way and the local intergalactic environment. With the advent of FPGA micro-chips for radio astronomical spec\-tro\-me\-ter \linebreak \citep{stanko2005,klein2009} such an approach became feasible. FPGA spectrometer offer a high degree of flexibility because the analog digital converter can be directly controlled and the FPGA-core allows to combine a broad range of spectral setups on bandwidth and number of spectral channels.
For EBHIS we use FPGA-spectrometers which were originally designed for the sub-mm range. An on-board quartz-oscillator triggers the analog digital converter to sample 100\,MHz bandwidth. The FPGA core provides 16.348 spectral channels with 6.1\,kHz separation.
The Allan-time of the FPGA spectrometer allows to integrate for a few hours without any deviation from the theoretical expectation. We tested the receiver and spectrometer setup using standard line calibration region \citep{kalb82} and showed that the spectrometer allows to fulfill \linebreak the EBHIS survey requirements on sensitivity, stability and dump speed \citep{winkel2010}.

\subsection{RFI Contamination}\label{rficlouds}
A severe limitation of the \hi data quality is introduced by radio frequency interference (RFI). The extragalactic portion of EBHIS covers the frequency range clearly out--side the radio astronomy protected band. 95\% of the RFI contamination within the 100\,MHz bandwidth is limited to numerous but single spectral channels \citep{floeer2010}. The remaining contamination can be characterized by broad  \newline band interference which occur only during a short period of time.

During the course of the EBHIS survey a third, new type of RFI contamination has been detected. Figure\,\ref{clouds} shows the grey-plot (spectral channels vs. time) the spectral data \linebreak which are degraded by ``cloudy'' structures. The brightness temperature variations of these RFI reach the 1\,K level and degrade the determination of the spectral baseline strongly.
Inspecting all recorded data it was feasible to identify observed fields contaminated by these RFI. We identified a strong azimuth dependence of the contamination. Between $230\degr$ and $360\degr$ the broad band interferences are so severe, that it is not feasible to overcome the bandpass degradation using a software algorithm. We need to re--observed those portions of the sky before the final data release. Applying a flexible scheduling scheme we now avoid this azimuth range (Fl\"oer et al. in prep.).

\subsection{Observational Setup}
To survey the whole sky we are aiming for multiple coverages of the total survey area. These multiple coverages will observe the same portion of the sky at different seasons, accordingly at different hour angles. This allows us to overcome seasonal irradiation from ground received by the near- and far side-lobes \citep{kalb1980,hartmann96}.
Multiple coverages are also an ideal tool to minimize the RFI contamination \citep{kalb2010}. Based on our experience with RFI, the contamination is a function of time and azimuth angle \citep{winkel2007}. The repeated observation of the same portion of the sky during different dates allows to overcome even long term RFI contamination.

The spectrometers record two polarizations every \linebreak 0.5\,sec. This fast storage of data yields a high data rate but allows us to sample the sky densely. Using a driving speed of $\frac{4\degr}{\rm min.}$ we sample the sky portion equivalent to area of the Effelsberg beam with about 30 individual measurements. We start at the low declination boundary and track along constant declination for $5\degr$ in right ascension. The individual scans in declination are separated by $8'$. After $2.5$\, min\-utes we cover the same portion of the sky with the off-set feeds. Short term RFI on a few minuted scale can be identified be comparing the central and the off-set beam signals \citep{floeer2010}. The individual fields of interest overlap in right ascension and declination by a single beam size of $9\arcmin$.
Because we sample the sky in equatorial coordinates, we scale the size of the fields of interest by 1/cos($\delta$).
Using this scanning speed we need about 65\,minutes for a single coverage of a field of interest including all overheads. The effective observing time yield an rms between 60 and 90\,mK per beam per spectral channel. Since the system temperature depends on elevation \citep{winkel2010}, multiple coverages will increase the overall quality of EBHIS.

\section{EBHIS Status and Data Quality}
Today EBHIS surveyed about 8000 square degrees (Fig\,\ref{ebhiscoverage}). Most of the already surveyed area is towards low declinations. Up to now, we did not cover the circumpolar sky of Effelsberg. This focus on low galactic latitudes allows us to evaluate the data quality in direct comparison to the Parkes GASS survey (see Sect.\,\ref{MWsky}). We expect to finish the first full sky coverage within about 900\,hours of necessary observing time mid 2011. Most of the observing time for EBHIS is scheduled for summer/autumn 2010 and summer 2011 because of the higher water vapor content during these seasons.

\citet{winkel2010} discussed the complex calibration process of the EBHIS data in great detail. However, despite the high demand on computation time we gain highly significant \hi data from their procedure with a calibration accuracy of better than 3\% across the whole bandpass.

The first EBHIS coverage of the full northern sky will offer a rms noise limit between 60\,mK and 90\,mK per beam and spectral channel depending strongly on elevation (see \citet{winkel2010}), allowing to determine the WNM column density with an accuracy of (${\rm FWHM = 20\,km\,s^{-1}}$) $N_{\rm HI} \leq 3.2\cdot 10^{18}\,{\rm cm^{-2}}$. For the extragalactic sky we planning to average the data in velocity to the equivalent velocity resolution of HIPASS to $18\,{\rm km\,s^{-1}}$ and we will reach a limiting sensitivity 13\,mJy. At the distance of the Virgo cluster we expect to detect a dwarf galaxy with a minimum \hi gas mass of about $M = 8\cdot 10^7\,{\rm M_\odot}$.

\subsection{The Milky Way Sky}\label{MWsky}
Figure\,\ref{ebhisivcsky} shows a single velocity channel of the EBHIS data at $v_{\rm LSR} = -47\,{\rm km\,s^{-1}}$. The rectangles enclose those portions of the sky, which we discuss in greater detail. In the middle, a rectangle marks the position of an IVC structure. \citet{winkel2010} showed in their Fig.\,14 a single velocity slice to demonstrate the agreement between both, GASS and EBHIS. Here, we focus on the detailed structure of the IVC gas in general and its relation to the Milky Way disk emission. 


\subsubsection{Example: Intermediate--Velocity Clouds}
Towards the whole field of interest we identified huge IVC structures (Fig.\,\ref{ebhisivcsky}) several tens of degrees in extent which form coherent structure also in the velocity domain. Starting around $v_{\rm LSR} = -90\,{\rm km\,s^{-1}}$ we find filamentary IVC structures covering the whole field of view. These filaments show up with tiniest details which are unresolved by the Effelsberg telescope (Fig.\,\ref{handballer} top panel). As in case of the HVC (see below) the IVCs fragment into individual clouds with angular sizes of about a single beam.
We selected to visualize the fragmentation of the IVC in individual clumps a radial velocity ($v_{\rm LSR} = -36\,{\rm km\,s^{-1}}$). Common to all compact HVCs and IVCs is that all the individual clouds are well separated from each other but embedded within an extended warm gas envelope. The nearly unresolved IVC clouds show up with emission lines of about $10\,{\rm km\,s^{-1}}$ \linebreak (FWHM), making them to transients between the CNM and the WNM. 
To quantify the IVC clumps we calculate the kinetic temperature form the \hi line profile width to $T_{\rm kin} = 21.8 \cdot \Delta v^2 = 2180\,{\rm K}$. The column density is $N_{\rm HI} = 3\cdot 10^{19}\,{\rm cm^{-2}}$. To evaluate the volume density we use two different approaches:
\begin{enumerate}
\item {\it observers approach:\/} Assuming that the IVCs are objects of the lower Galactic halo, we adopt a vertical distance from the Galactic plane of about 2\,kpc. The distance is about 5.5\,kpc yielding for an unresolved structure a linear cloud extent of 14\,pc. We derive with these numbers a volume density of $n = 0.7\,{\rm cm^{-3}}$ and a pressure of ${\rm log}(P/{\rm k}) = 3.1\,{\rm K\,cm^{-3}}$.
\item {\it theoretical approach:\/} Using the size-linewidth relation of \citet{wolfire2003} $l_{\rm p} = 0.3\cdot \left(\frac{T_{\rm kin}}{\rm [100\,K]}\right)^\frac{3}{2}$ we calculate the linear size of the clump to 30\,pc. This yields a factor of two lower volume density and hence pressure.
\end{enumerate}
Using \citet{wolfire1995b} their Fig.\,1(a) we deduce, that the IVC clumps are close to pressure equilibrium with the surrounding medium. The size--line--width relation determines the maximum size of the clump where pressure confinement and turbulence are of equal importance. We find, that the individual IVC clumps are close to equilibrium with the gaseous envelope.
But the linear alignment, the periodic appearance of the clumps within the IVC filament (see Fig.\,\ref{handballer} might be considered as indications for Ray\-leigh--Tay\-lor instabilities. The \hi mass of a single IVC clump is only about $M = 34\,M_\odot$ much below the mass of halo clumps studied i.e. by \citep{ford2010}.

A lower IVC velocity boundary (towards the Milky Way standard rotation) can not be identified.
Continuous connections between the IVC and Milky Way disk are present across the whole field of interest. As complex as the IVC gas structure is the Milky Way gas structure towards this portion of the sky, as demonstrated by the individual \hi spectra. The line shape of the Milky Way emission varies appreciable on angular scales below $1\degr$. Figure\,\ref{handballer} (bottom panel) shows two spectra separated by $75\arcmin$. The IVC emission line is visible around $v_{\rm LSR} = -38\,{\rm km\,s^{-1}}$. The line emission of the Milky Way galaxy shows strong changes in intensity and line shape on this angular scale. 

A more detailed investigation of the IVCs appears to be promising to study the impact of high altitude gas on the structure of the Milky Way galaxy on global scales.


\subsubsection{Example: High--Velocity Clouds}
Within the field of interest are HVC complexes GCN/GCP are located (Fig.\,\ref{ebhisivcsky} marked by the rightmost rectangle). We focus in the following on HVC\,043.7-12.5-318. Figure\,\ref{GCNUCHVC} \linebreak shows the brightness temperature distribution of the \linebreak UCHVC around $v_{\rm LSR} = 330\,{\rm km\,s^{-1}}$. This HVC has an angular extent of about $0.5\degr$. It fragments into two cores which are embedded within an envelope of warm gas. These two cores are Marked by the circles in Fig.\,\ref{GCNUCHVC}, both not resolved by the Effelsberg telescope.
The right hand panel of Fig.\,\ref{GCNUCHVC} shows the \hi spectra towards both cores. The column densities of both clumps are about the same with $N_{\rm HI} \simeq 3\cdot 10^{19}\,{\rm cm^{-2}}$ while the velocities are different by $\Delta v_{\rm LSR} = 55\,{\rm km\,s^{-1}}$!
This complex velocity structure and the small separation of both \hi cores is identified in the LAB data as a so--called head--tail structure \citep{bruens2000}. The EBHIS data however discloses, that both cores show up with rather broad emission lines (FWHM = $30\,{\rm km\,s^{-1}}$) implying $T_{\rm kin} \sim 20.000$\,K.
Assuming 10\,kpc \citep{wakker2001} as an distance estimate, we derive a volume density of \linebreak $n_{\rm HI} \simeq 0.4\,{\rm cm^{-3}}$. This implies a pressure of the neutrals gas ${\rm log}(P/{\rm k}) = 3.9\,{\rm K\,cm^{-3}}$. \citet{wolfire1995b} show in their Fig.\,3 (panel (e) for primordial gas composition) that for a stable cloud in equilibrium with the ambient halo gas the distance is in the range $7.5 \leq z[{\rm kpc}] \leq 10$. This is inconsistent with the low galactic latitude of the cloud of interest, so our initial estimate is much too low or the hypothesis of a pressure equilibrium is incorrect.

To prove the stability hypothesis we use our temperature estimate and calculate the sound velocity within both cores to $c_{\rm Sound} = 14.5\,{\rm km\,s^{-1}}$. This implies that the sound speed is about a factor of four lower than the radial velocity difference of both cores in the common envelope. This suggests the presence of strong shocks across the whole \hi cloud. The sound crossing time is about $t_{\rm cross} \sim 8.5\cdot 10^5 \times D[{\rm 10\,kpc}]\,{\rm years}$, accordingly we see most likely un-relaxed physical structures. \citet{wolfire1995a} Fig.\,3 \newline panel (c) implies for a stable cloud an up to 10\% ionization of the gas. So, we may speculate that we detect only the neutral fraction of a gaseous cloud which is significantly ionized \citep{shull2009}. If this scenario is correct, the neutral gas is just the tracer for enhanced volume densities which cools more efficient that the ionized medium which carries the momentum.

An impressive demonstration how misleading angular under-sampled \hi can be is shown in Fig.\,\ref{BWUCHVC}. Shown here is the brightness temperature distribution of LAB (grey-scale) and EBHIS (contour lines) of a portion of HVC complex GCP. The prominent HVC at $l = 58\degr$ and $b = -21\degr$ shows in the LAB a complete different structure compared to \newline EBHIS. The positions of the temperature maxima are not well recovered by LAB. The UCHVC at $l = 61\degr$ and $b = -19\degr$ is undetectable with the resolving power and sensitivity of LAB. This leads to the conclusion, that most of the CNM/WNM small-scale structure is towards the northern sky only accidentally detected towards the northern sky. A comprehensive view of HVC complex GCN/GCP will be presented by Winkel et al. (in prep.).

The higher angular resolution of the EBHIS data towards HVCs and IVCs suggest the existence of \hi structure close to the angular resolution limit of the 100-m dish. Fragments rather than smooth \hi gradients become detectable. The 100-m beam obviously couples better than the LAB data on the relevant angular scales of the WNM gas far away from the Milky Way disk. The smooth head--tail structures \citep{bruens2000} might be partly interpreted as smooth and continuous velocity gradients due to the insufficient angular resolution of the previous all--sky surveys.



\subsection{The Extragalactic Sky}
We evaluated the performance of the receiving system by studying the Virgo galaxy cluster. Our aim is to detect with\-in the first EBHIS coverage dwarf galaxies with a mass of $1.2\cdot 10^8\,{\rm M_\odot}$ at the distance of the Virgo cluster. Even with FPGA spectrometers and eight bit data sampling the continuum radiation of Virgo A saturates the spectrometer. Accordingly we need a very careful data analysis to reach our sensitivity goal. Applying the data reduction chain of \citet{winkel2010} we reach this expected detection limit. Deep EBHIS test fields towards the Virgo cluster show that in a targeted search for dwarf galaxies we reach an \hi mass limit of $3\cdot 10^7\,{\rm M_\odot}$ with an effective on--source time of 7\,minutes. This detection limit demonstrate the sensitivity of the EBHIS data and is promising for the deep EBHIS SDSS survey. The detailed analysis of the Virgo data set and the discussion of the systematic limitations will be presented by Fl\"oer et al. (in prep.).

Towards the remaining sky we already detected numerous galaxies. Figure\,\ref{galaxymosaic} shows the beam--averaged \hi spectra of NGC\,4244 and NGC\,4395. These galaxies are covered by our northern galactic pole area. Unfortunately most of these data is affected by strong broad band RFI (see Sect.\,\ref{rficlouds}) and we reach currently only an rms limit of about 150\,mK. However, these galaxies are luminous in \hi radiation and the limiting column density integrated across $100\,{\rm km\,s^{-1}}$ is about $N_{\rm HI} \leq 1.4\cdot 10^{19}\,{\rm cm^{-2}}$. A sufficient basis for short-spacing corrections of radio interferometric data. The \hi mass detection limit of the EBHIS first all--sky coverage will be 
\begin{equation}
M \geq 6200\,{\rm M_\odot} \left(\frac{\Delta v}{\rm [km\,s^{-1}]}\right)\left(\frac{D^2}{\rm [Mpc]}\right)
\end{equation}



\subsection{The Continuum Radiation}
The FPGA spectrometers offer 8-bit digital sampling. Accordingly, the spectral data allows us to evaluate the absolute continuum level. Using the data reduction chain of \citet{winkel2010} the bandpass shape is precisely modeled. This allows to evaluate the total value of the underlying continuum level. As a ``spin--off'' product, EBHIS contains valuable information on the position and intensity of radio continuum sources across the 100\,MHz band. In Fig.\,\ref{contsources} we show the radio continuum source distribution towards a 30 square degree area. The standard SEx\-trac\-tor \linebreak \citep{Sextractor} identifies 30 sources above a \linebreak threshold of 0.15\,mK. Elongated and extended sources were rejected from further analysis to compare the unresolved EBHIS sources with NVSS data \citep{condon1998}. The remaining list of 24 sources we searched for NVSS counterparts. All sources are real detections. Figure\,\ref{radcont} shows the correlation of EBHIS $T_{\rm B}$ versus NVSS flux. The uncertainties of the EBHIS fluxes are evaluated by SExtractor. We find a very well linearity between EBHIS and NVSS fluxes! Finally we evaluated the positional accuracy of the EBHIS data with the NVSS positions. 74\% of all continuum sources are encircled within a radius of $2\arcmin$. Using a driving speed of $4\arcmin$ per second, $2\arcmin$ corresponds to the angle between to consecutive spectral dumps. The analyzed data cube offers a pixel size of $3\arcmin \times 3\arcmin$. According to Fig.\,\ref{contpositions} the EBHIS positions match the NVSS positions very well. This confirms that the data reduction process is also properly recovering the positional information. Up to now, we covered only a minor fraction of the Milky Way disk emission. Towards the immediate vicinity of the Galactic disk we already identified also diffuse radio continuum emission. As a real ``spin--off'' product, EBHIS will yield a northern sky radio continuum survey and might substitute the Stockert radio continuum survey \citep{reich1982}. We expect to reach on average the 100\,mJy level towards the northern sky.  

\section{Summary}
The Effelsberg--Bonn HI survey covers the whole northern sky north of Dec(2000) $= -5\degr$. The first coverage will be completed in mid 2011. Due to the large bandwidth of about 100\,MHz and the FPGA spectrometers with 16.384 spectral channels, we perform a Milky Way and a local volume \hi survey in parallel. For the Milky Way \hi distribution we expect to reach all-over the sky an rms limit of about 90\,mK at 9\,arcmin angular and $1.3\,{\rm km\,s^{-1}}$ velocity resolution. For the extragalactic sky EBHIS will detect all $M(HI) \geq 1.2\cdot 10^8\,{\rm M_\odot}$ galaxies at the distance of the Virgo cluster. Part of the EBHIS project is a deep \hi survey of the SDSS area. With 10\,minutes of integration time per position we improve the mass detection rate by a factor of four. First performance tests by Fl\"oer et al (in prep.) show the feasibility of the project. Smoothing the spectral resolution to the velocity resolution of HIPASS already the first EBHIS coverage of the northern sky will complete on nearly the same sensitivity limit the HIPASS survey. EBHIS and GASS show up with comparable spectral and angular resolution as well as sensitivity. The combination of both surveys will improve our knowledge of the Milky Way column density distribution by nearly an order of magnitude. Here, full angular sampling is the driver for precise astrophysical information, necessary in particular for high energy astrophysics. As a ``spin--off'' product, EBHIS contains valuable information on the continuum radiation of the sky. EBHIS demonstrates the power of FPGA spectrometer for radio astronomy, offering spectral line and continuum surveys in a single run.

\acknowledgements We thank our referee F.J. Lockman for his detailed comments on the manuscript. We also thank the Deutsche Forschungsgemeinschaft (DFG) for the project funding under \newline grant KE757/7-1. Based on observations with the 100-m telescope of the MPIfR (Max-Planck-Institut f\"ur Radioastronomie) at Effelsberg

\bibliographystyle{aa}
\bibliography{references}

\clearpage

\begin{figure*}
\centerline{
\includegraphics[scale=0.5]{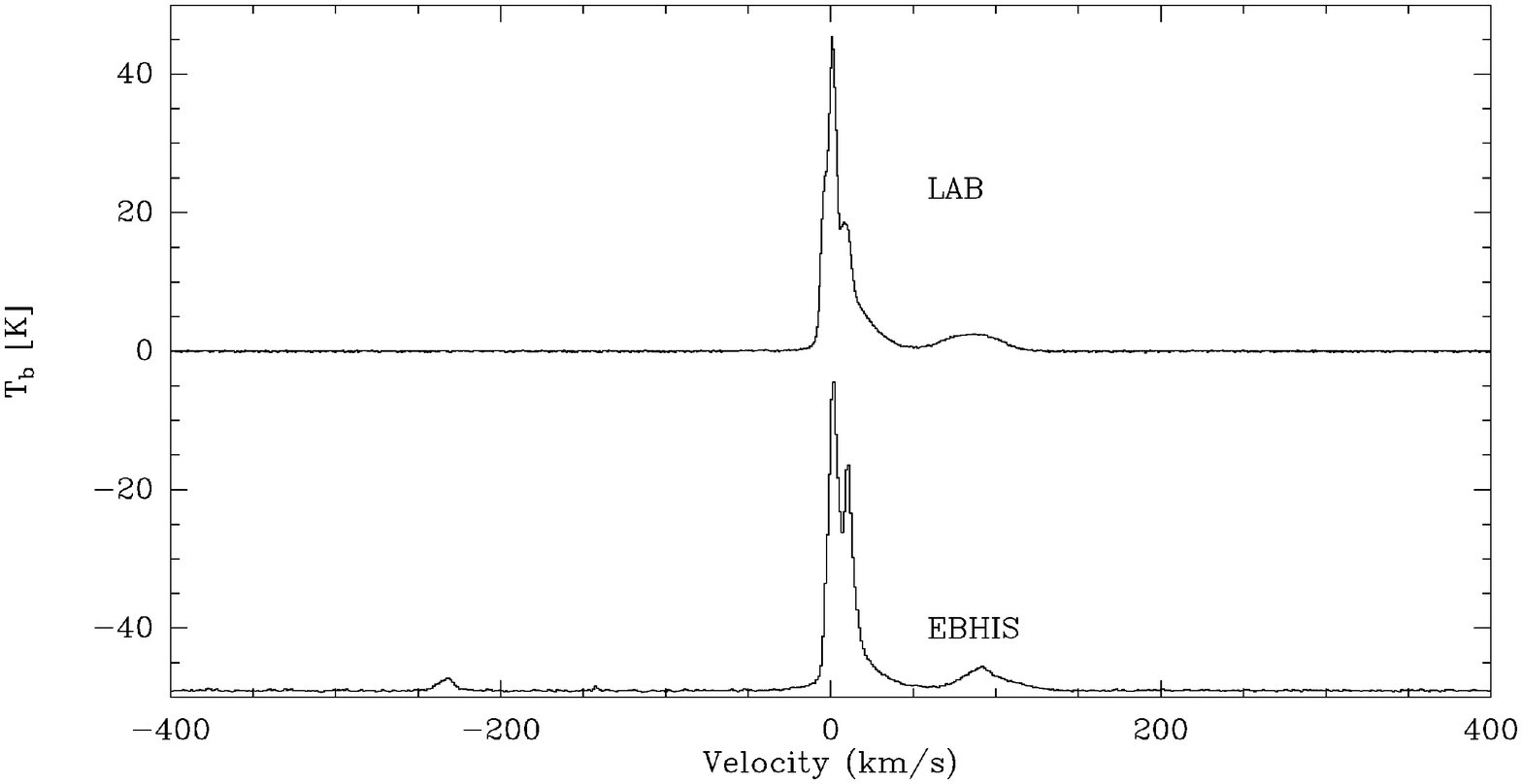}
}
\centerline{
\includegraphics[scale=0.5]{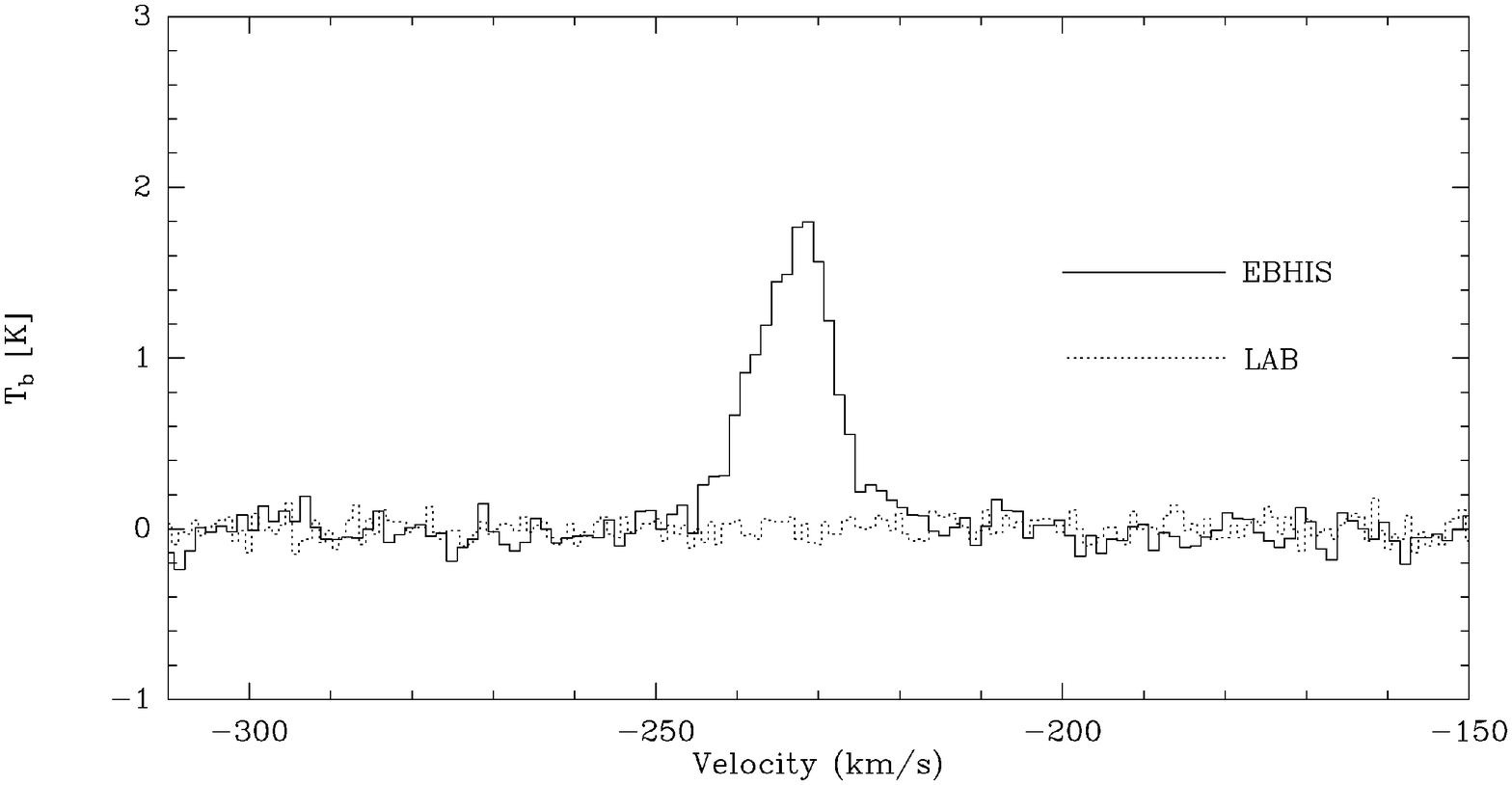}
}
\caption{{\bf Top:} Towards the HVC complex GCN EBHIS disclosed the arrangement of a number of HVCs which are hardly resolved by the Effelsberg telescope. EBHIS detected emission of HVC\,041-14-230 while the LAB survey \citep{kalb2005} appears to miss the emission line. Obviously the higher angular resolution of the Effelsberg dish discloses additional small scale structures within the Milky Way \hi distribution. The EBHIS line profile shows the presence of cold gas, traceable by the narrow \hi lines. {\bf Bottom:} Detail of the \hi emission of HVC\,041-14-230. The solid line shows the EBHIS data while the dashed line shows the LAB data \citep{kalb2005}. The difference in the  beam--filling would yield a 125\,mK signal of the cloud in the LAB data. The non-detection is because of the angular under-sampling of the sky.\label{labvsebhis}}
\end{figure*}
\clearpage
\begin{figure*}
\centerline{
\includegraphics[scale=1.0]{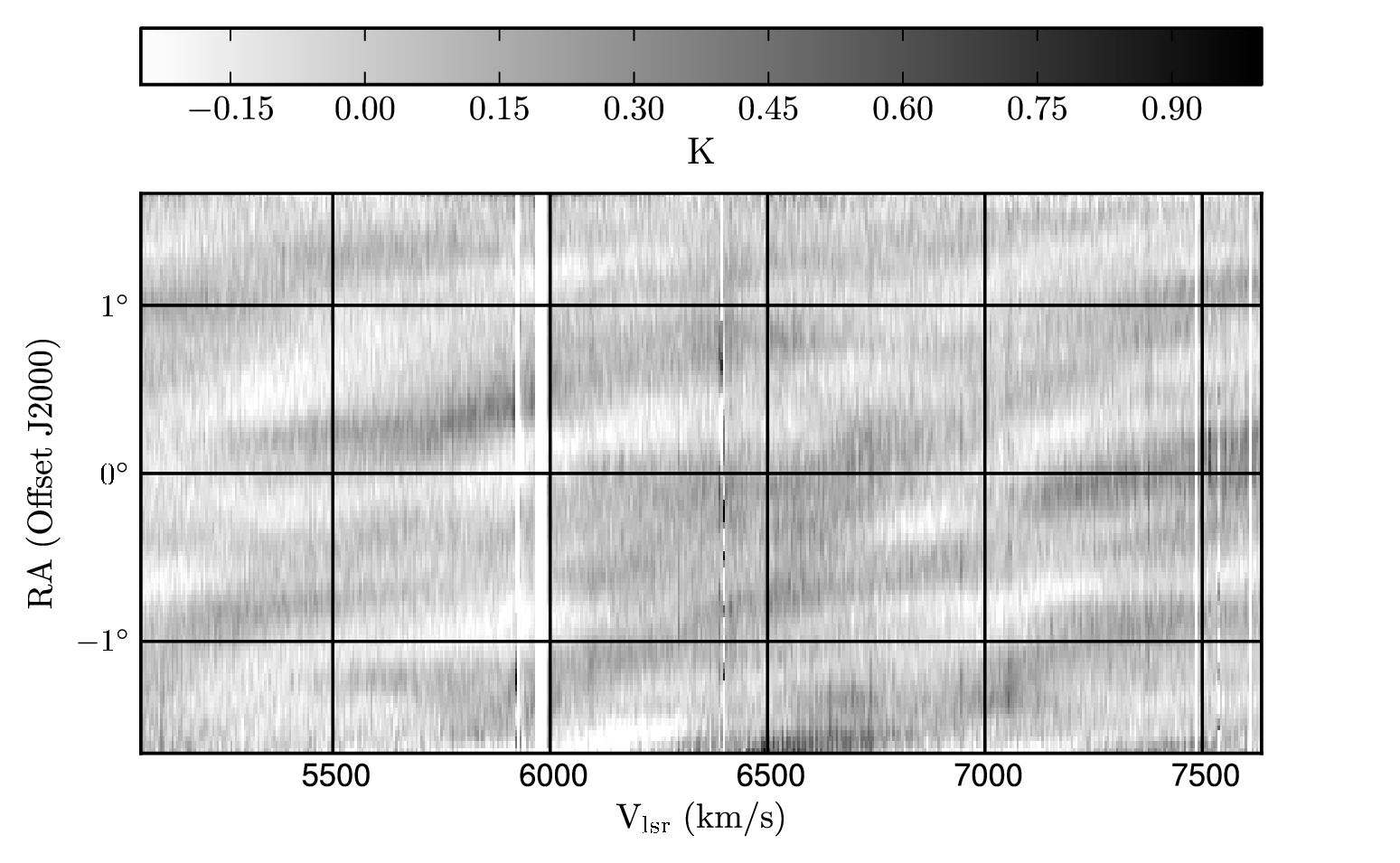}
}
\caption{EBHIS is affected by RFI signals. Up to 95\% of the RFI events can be identified and mitigated by the data reduction \citep{floeer2010}. However within a certain azimuth range between $230\degr$ and $360\degr$ the data is affected by strong broad band RFI signals. They show up with up to 1\,K brightness temperature across the whole bandpass. The figure show a ``grey-plot'' consisting of radial velocity versus right ascension. Each horizontal line represents a single spectral dump. The broad band interference varies in frequency with observing time but the pattern remains constant. The change in frequency might be partly due to the correction of the local standard of rest velocity frame. Data affected by these RFI events suffer from significant uncertainties in the baseline. Accordingly, the final spectrum shows up with an average rms of about 150\,mK and will be re-observed until summer 2011.\label{clouds}}
\end{figure*}
\clearpage
\begin{figure*}
\centerline{
\includegraphics[scale=1.1]{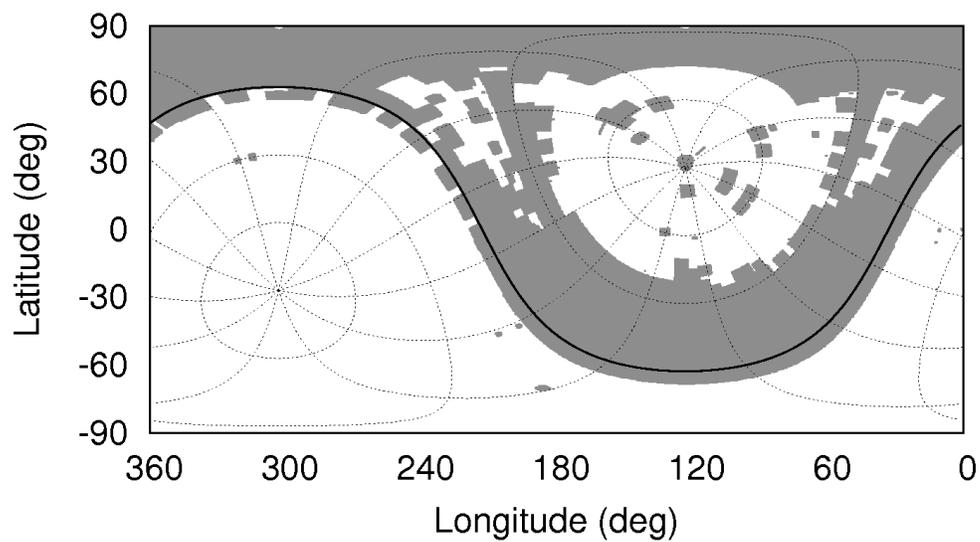}
}
\caption{Coverage of the EBHIS full northern \hi survey in autumn 2010 presented in galactic coordinates. The inhomogeneous coverage is due to two track breaks of the Effelsberg telescope in 2009. We concentrate on the low declination area first. The first coverage of the whole sky is planned to be completed in summer 2011. The rms per velocity channel ($\Delta v_{\rm LSR} = 1.3\,{\rm km\,s^{-1}}$) is on average between 60 and 90\,mK. \label{ebhiscoverage}}
\end{figure*}
\clearpage
\begin{figure*}
\centerline{
\includegraphics[scale=0.17]{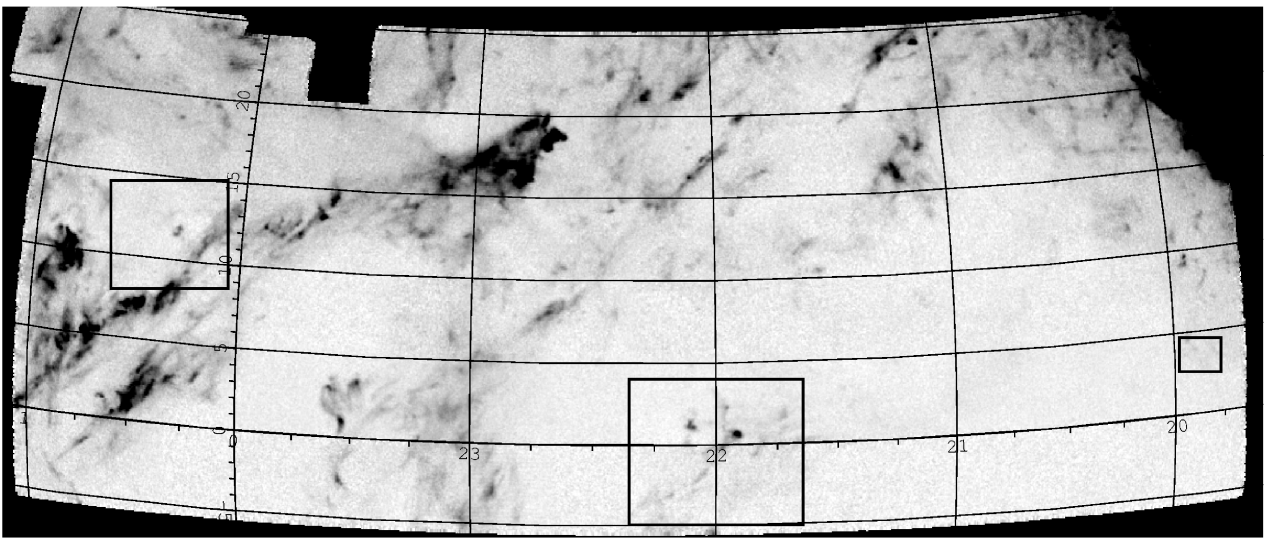}
}
\caption{Single velocity channel map ($v_{\rm LSR} = 47\,{\rm km\,s^{-1}}$) of a large EBHIS area. On a linear scale from 0.2\,K to 11\,K black denotes high brightness temperatures while white mark low ones. A wealth of large scale IVC filaments are visible which disclose structures down to the angular resolution level of the 100-m dish. Marked by rectangles from the left to the right are the positions of the radio continuum evaluation area (Fig.\ref{radcont}), the IVC clump area (Fig.\,\ref{handballer}) and the UCHVC belonging to HVC complex GCN (Fig.\ref{GCNUCHVC}). \label{ebhisivcsky}}
\end{figure*}
\clearpage
\begin{figure*}
\centerline{
\includegraphics[scale=0.5]{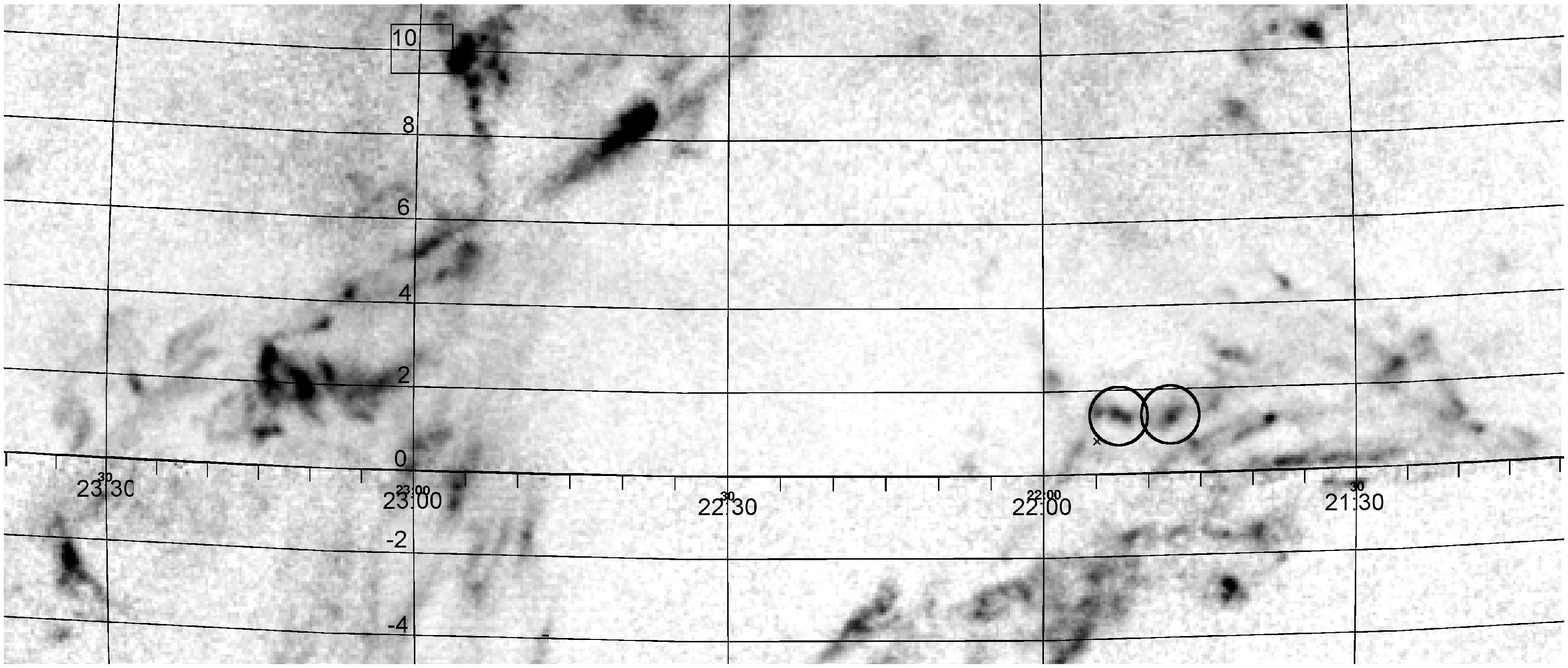}
}
\centerline{
\includegraphics[scale=0.5]{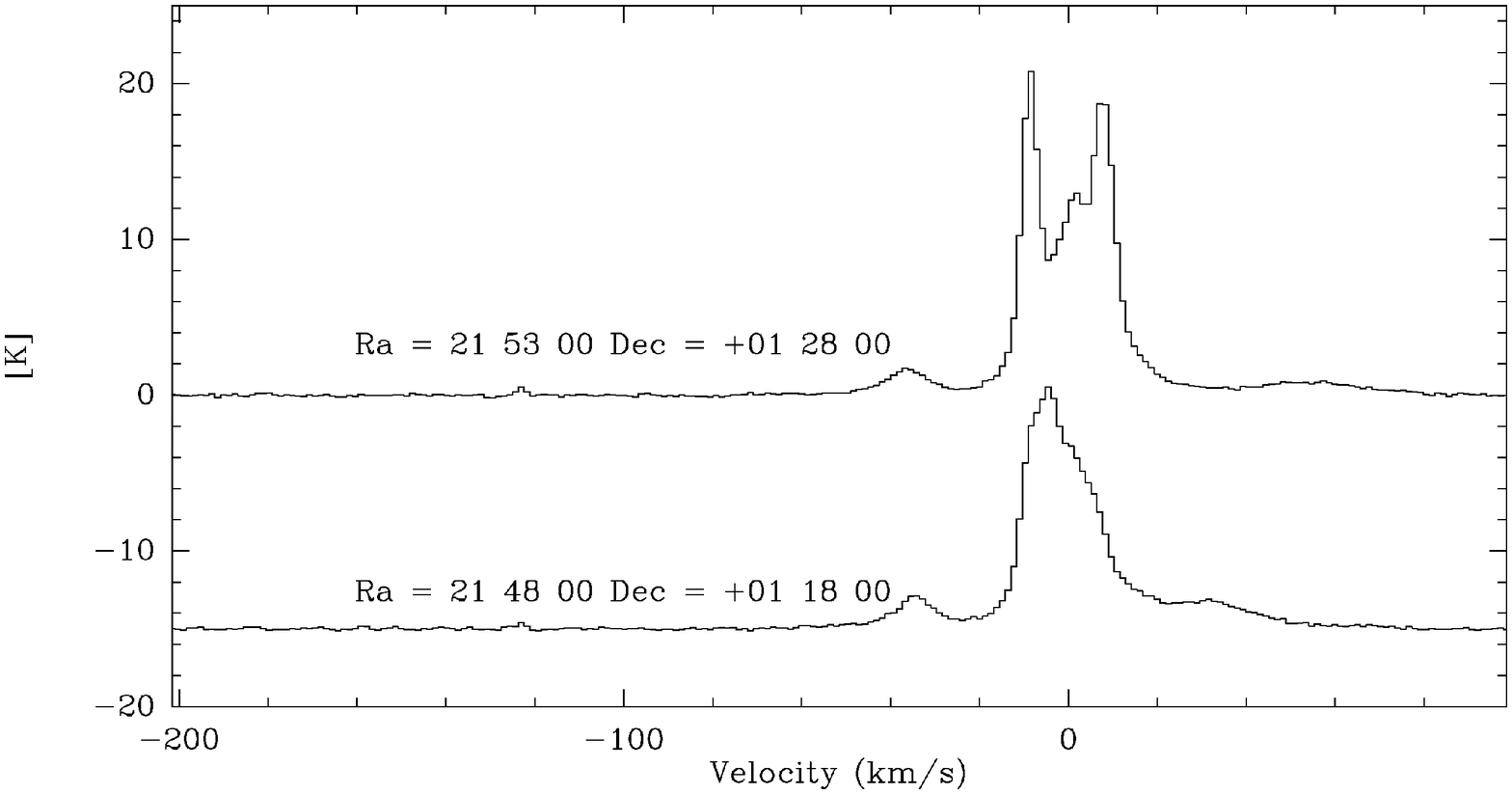}
}
\caption{{\bf Top:} EBHIS data of a single velocity channel ($v_{\rm LSR} \simeq -34\,{\rm km\,s^{-1}}$) towards the IVC structures in the overlap region with the GASS survey \citep{naomi2009, kalb2010}. Black denotes high, white low brightness temperatures ($ 0 \leq T[{\rm K}] \leq 3$) on a linear scale. Several IVC clumps are visible. All embedded within an envelope of \hi gas. The black circles mark the location of the IVC clumps. The center positions were used to extract the \hi spectra shown below. {\bf Bottom:} Selection of two EBHIS spectra towards IVC clumps separated by $75\arcmin$ angular separation (center positions are marked by circles in top panel). Remarkably different is the Milky Way \hi distribution, suggesting small-scale structure not only in the IVC but also in the low-velocity gas towards high galactic latitudes.\label{handballer} 
}
\end{figure*}
\clearpage
\begin{figure*}
\centerline{
\includegraphics[scale=0.4]{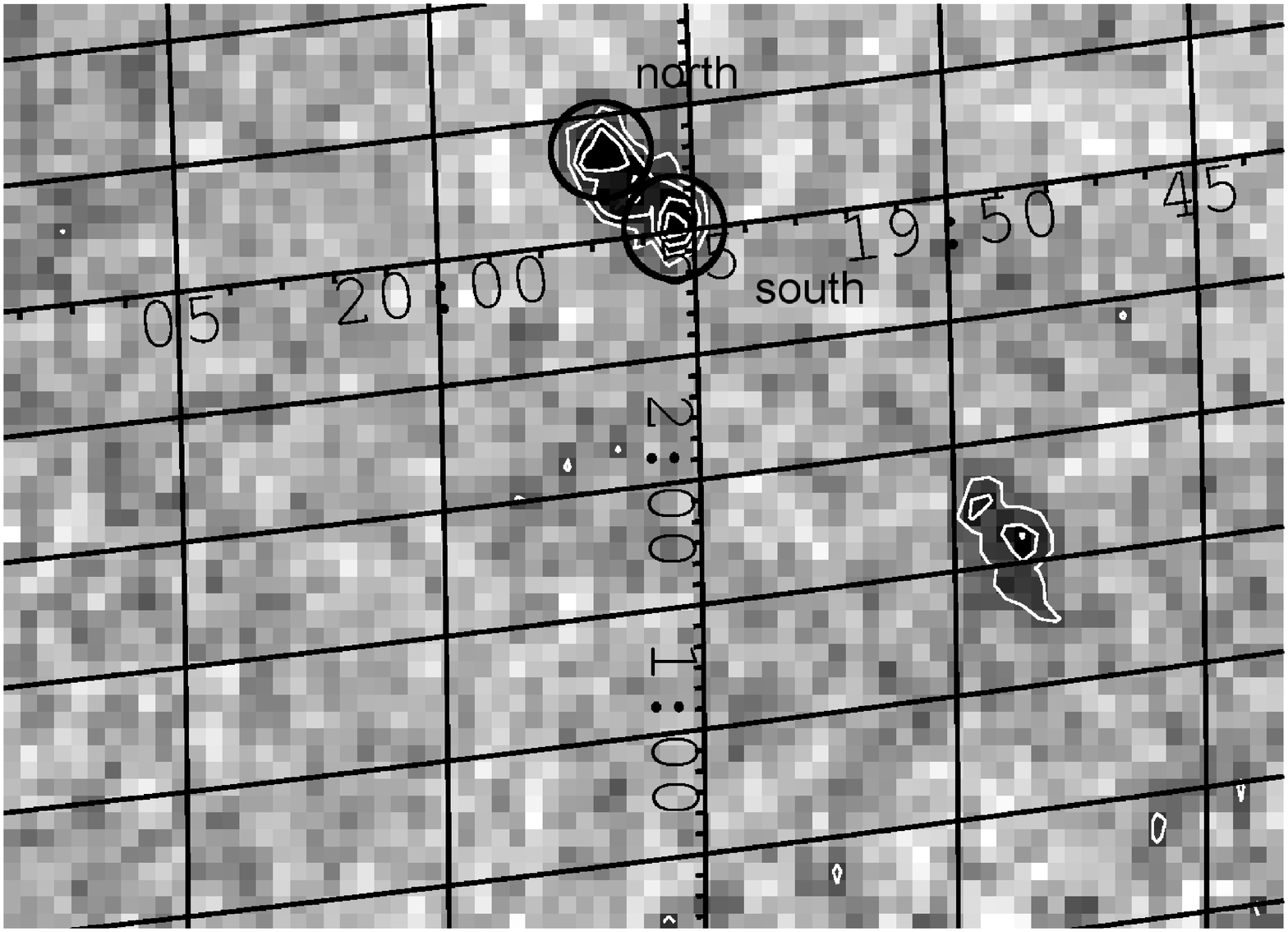}
}
\vspace{0.5cm}
\centerline{
\includegraphics[scale=0.5]{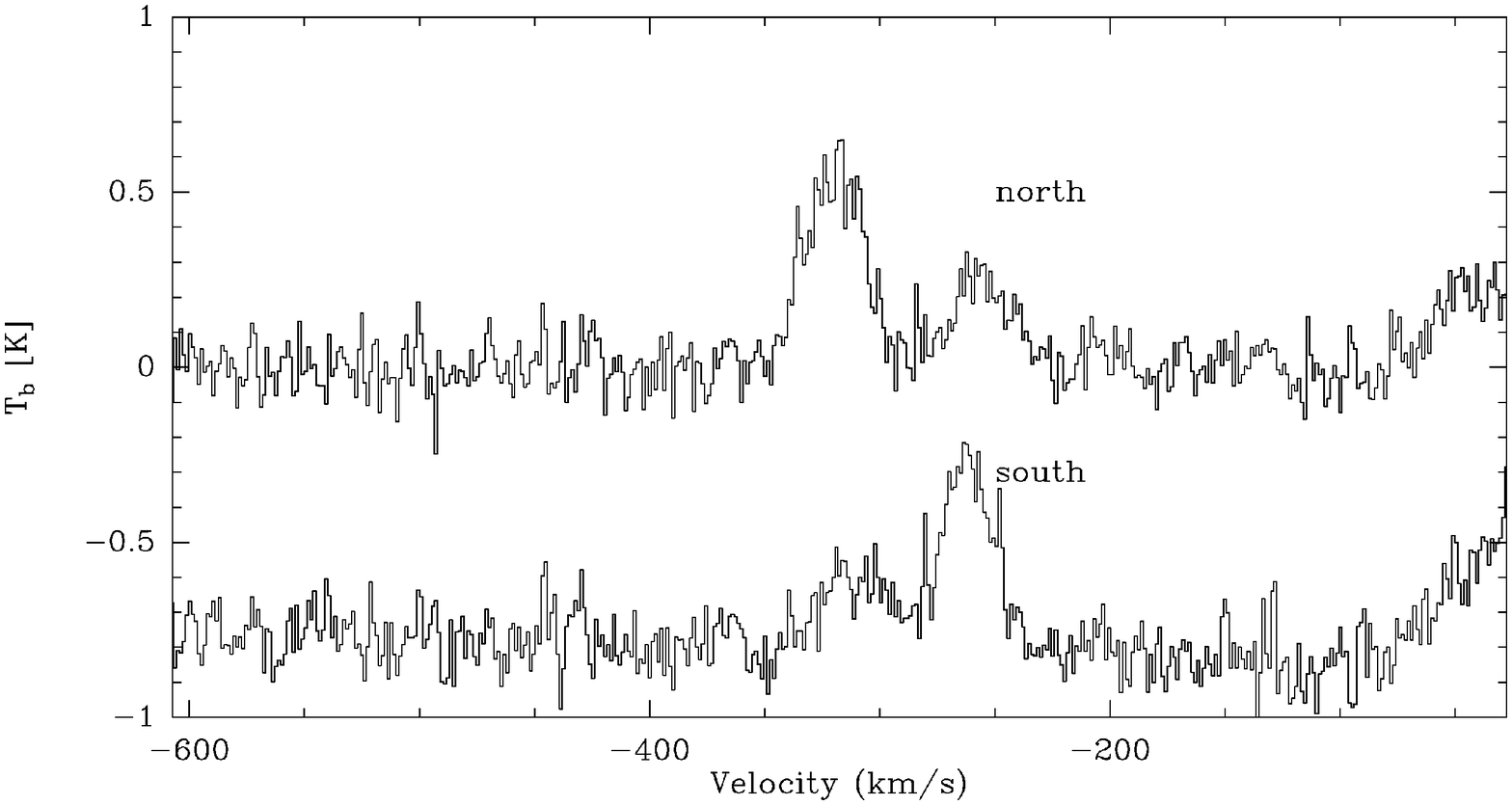}
}
\caption{{\bf Top:} The GCN HVC\,043.7-12.5-318 shows up with a very complex velocity structure. Enclosed in an \hi envelope we identify two column density maxima, each with $N_{\rm HI} \simeq 3.0\cdot 10^{18}\,{\rm km\,s^{-1}}$. The northern column density maximum shows up with a velocity of $v_{\rm LSR} = 318\,{\rm km\,s^{-1}}$ while the southern is much lower $v_{\rm LSR} = 263\,{\rm km\,s^{-1}}$. Assuming a primordial gas composition, the radial velocity difference is about $4 \times c_{\rm sound}$ with $c_{\rm sound} = 15\,{\rm km\,s^{-1}}$. {\bf Bottom:} \hi spectra of HVC\,043.7-12.5-318 towards the northern and southern core, respectively. The circle in the left hand panel mark the positions of the data extraction. \label{GCNUCHVC}}.
\end{figure*}
\clearpage
\begin{figure*}
\includegraphics[scale=0.8]{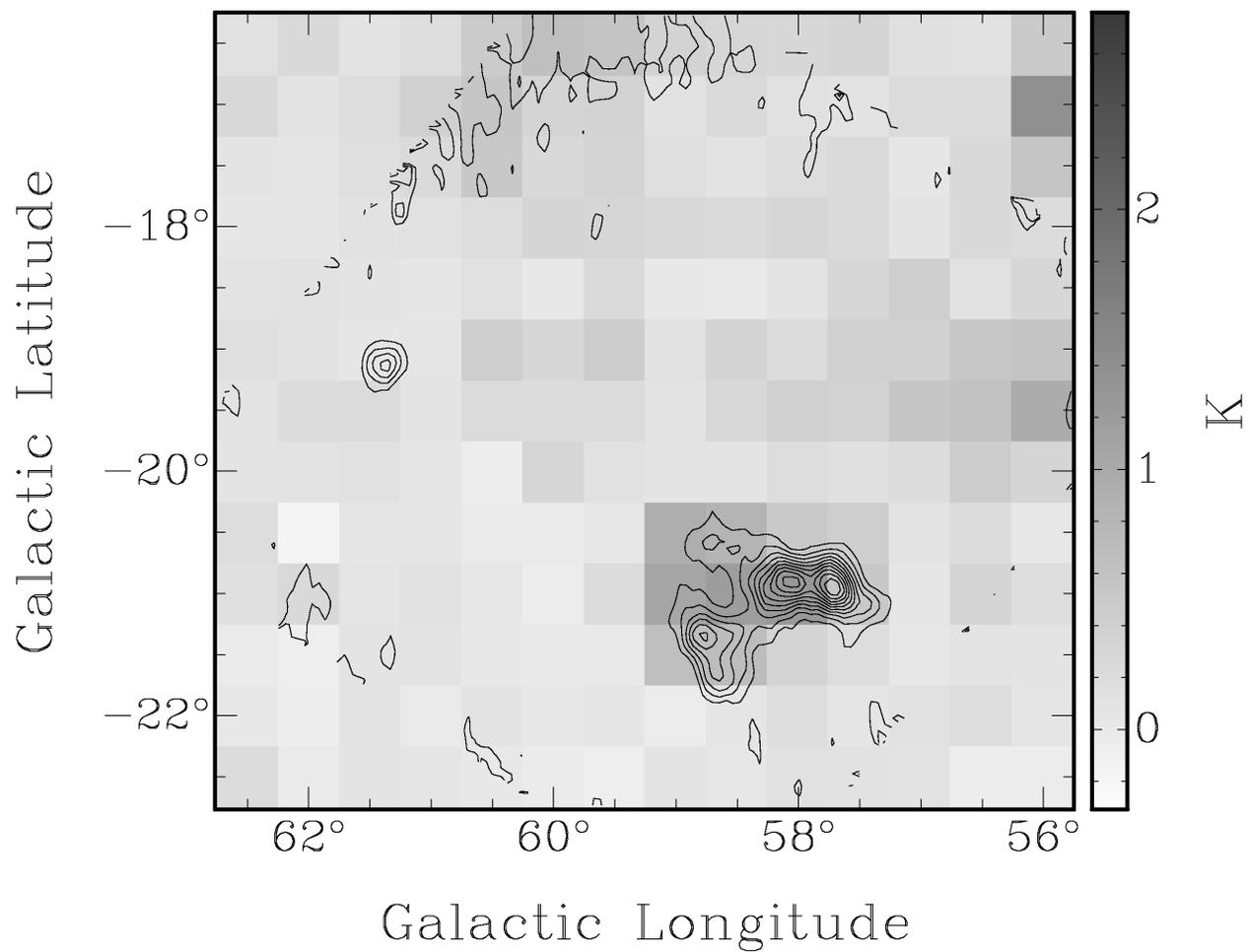}
\caption{Comparison on the detection limit and structure of an UCHVCs between LAB and EBHIS. The grey-scale map shows the LAB \hi brightness temperature distribution of an area towards HVC complex GCP while the contour lines mark the EBHIS data. There is an obvious positional offset between the maxima of the temperature distributions, also the morphology of the HVC appears different in both data sets. The UCHVC at $l = 61\degr$ and $b = -19\degr$ is not detected at all by LAB, here the difference in angular resolution and sensitivity lead to the non-detection by LAB. \label{BWUCHVC}}
\end{figure*}
\clearpage
\begin{figure*}
\centerline{
\includegraphics[scale=0.5]{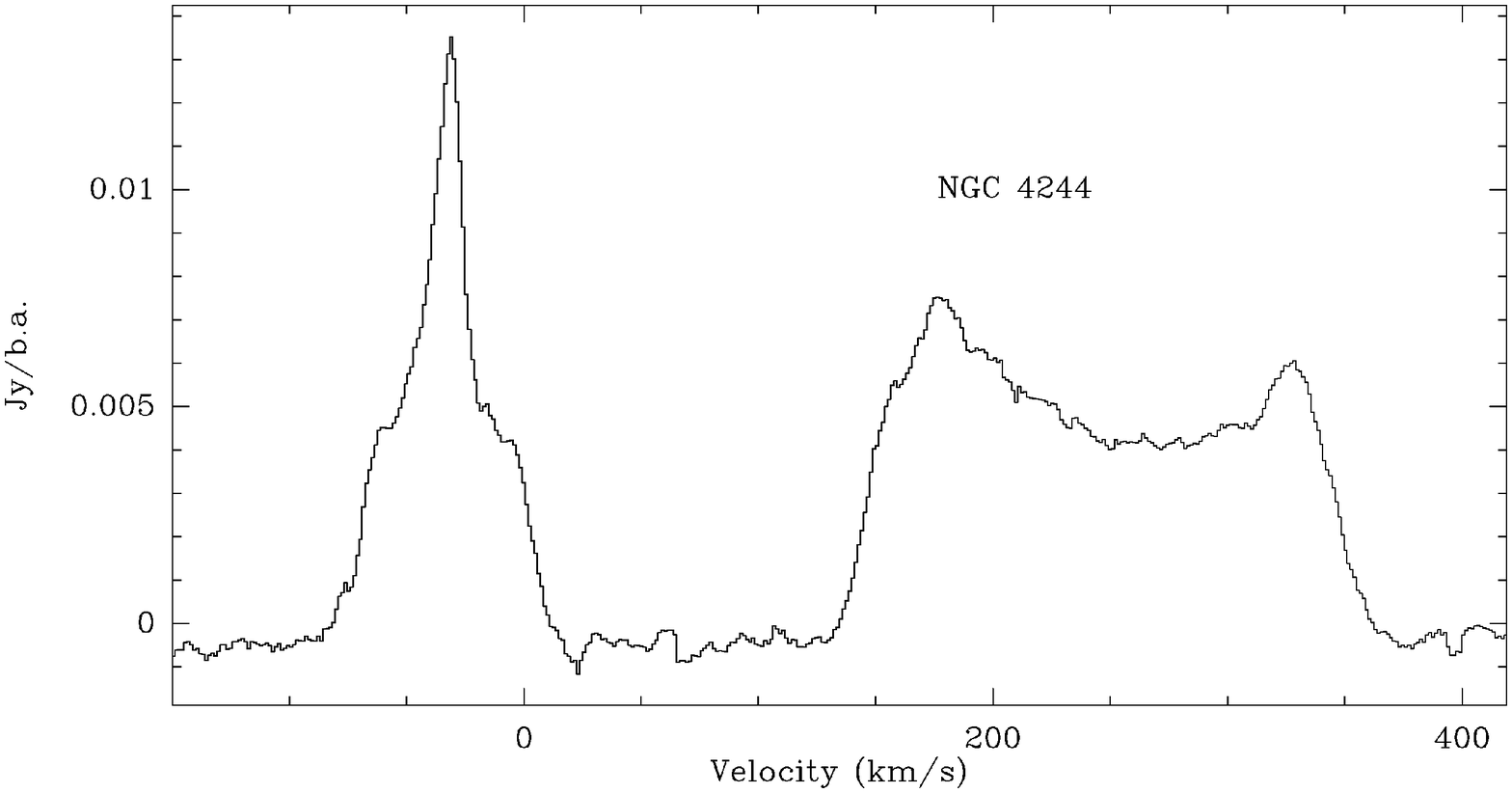}
}
\vspace{1.0cm}
\centerline{
\includegraphics[scale=0.5]{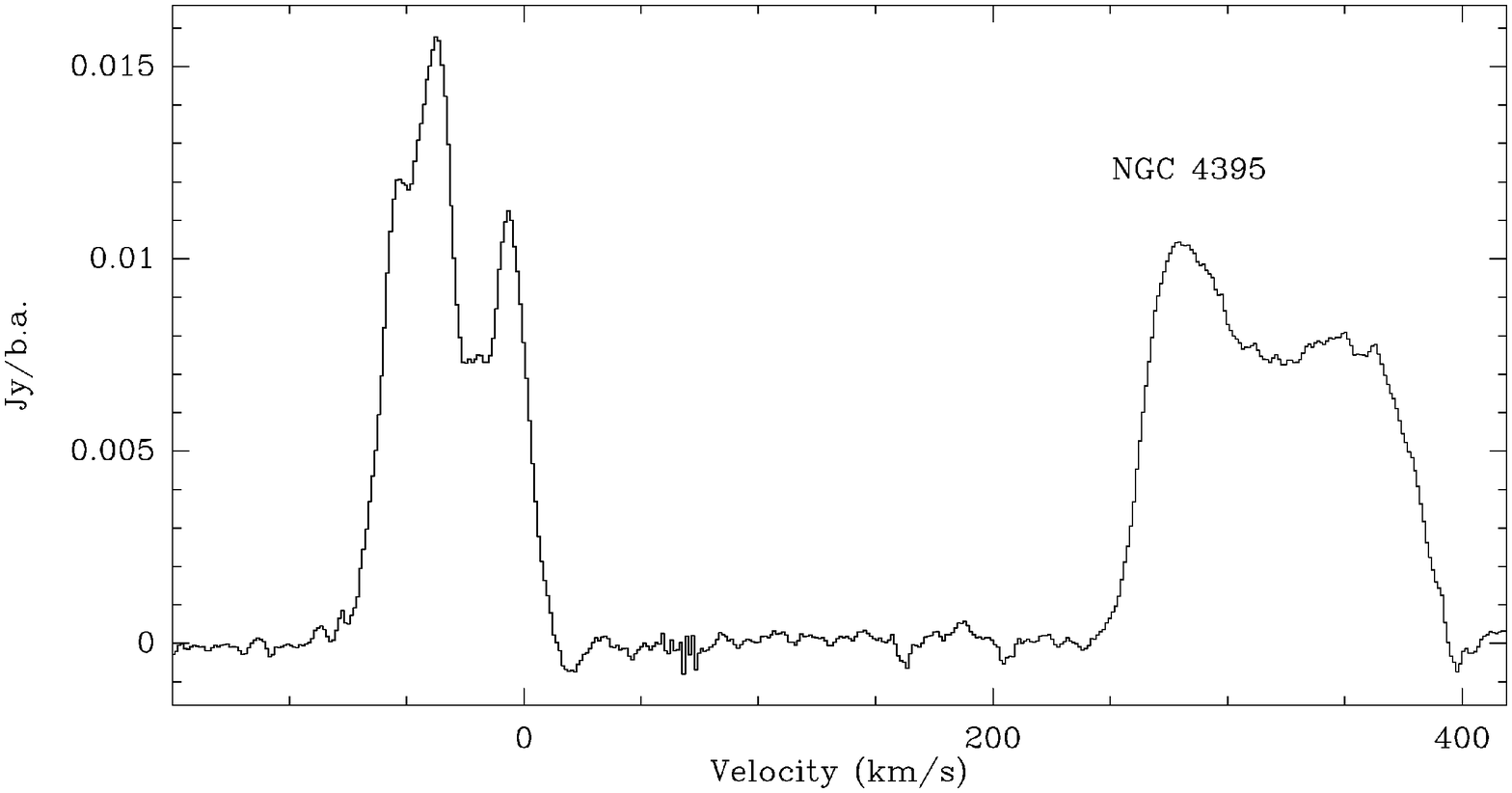}
}
\caption{To illustrate the quality of the extragalactic EBHIS data we show with full velocity resolution of $\Delta v_{\rm LSR} = 1.3\,{\rm km\,s^{-1}}$ the rotation curves of two galaxies. {\bf Top:} Rotation curve of NGC\,4244. {\bf Bottom:} Rotation curve von NGC\,4395. Residual RFI contamination is visible, because the data is affected by the strong RFI signals shown in Fig\,\ref{clouds}. Until mid 2011, the affected areas will be re--observed. \label{galaxymosaic}}.
\end{figure*}
\clearpage
\begin{figure*}
\centerline{
\includegraphics[scale=2.0]{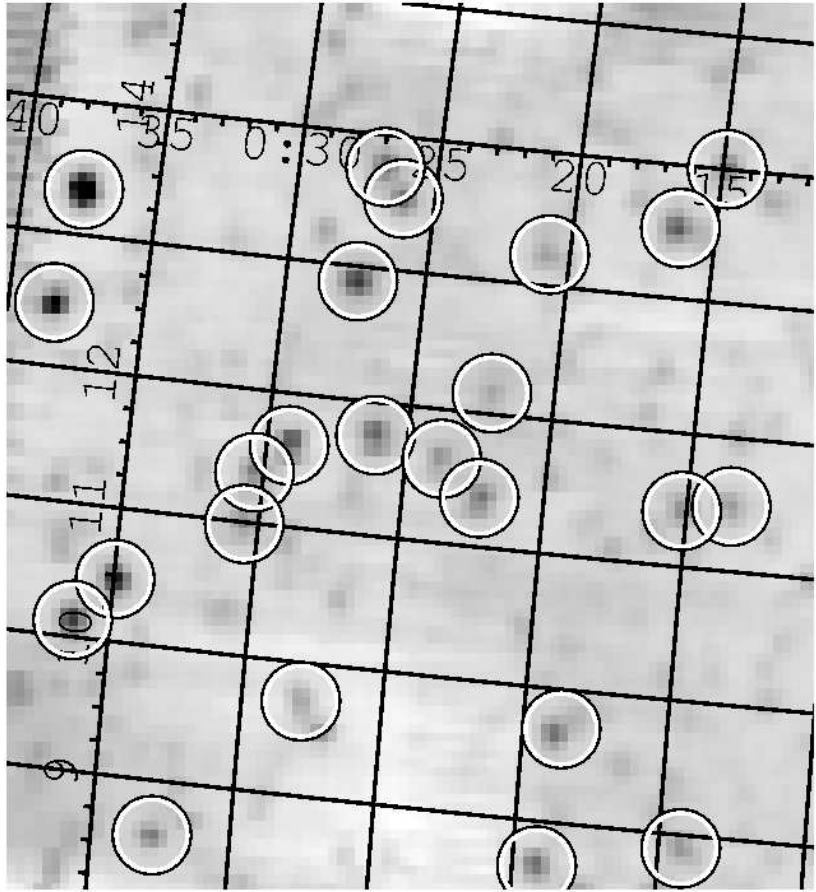}
}
\caption{EBHIS radio continuum information towards an arbitrarily chosen $5\degr \times 5\degr$ field of interest. Black denotes high fluxes while white denote low values. The circle mark the identified continuum sources using the standard source detection via SExtractor \citep{Sextractor}. Using a Gaussian probability distribution covering $3 \times 3$ pixels ($12\arcmin \times 12\arcmin$ area), the sources were identified. All extended sources were rejected from further evaluation. In total 24 source were detected and successfully identified using NVSS.\label{contsources}}
\end{figure*}
\clearpage
\begin{figure*}
\centerline{
\includegraphics[scale=0.6]{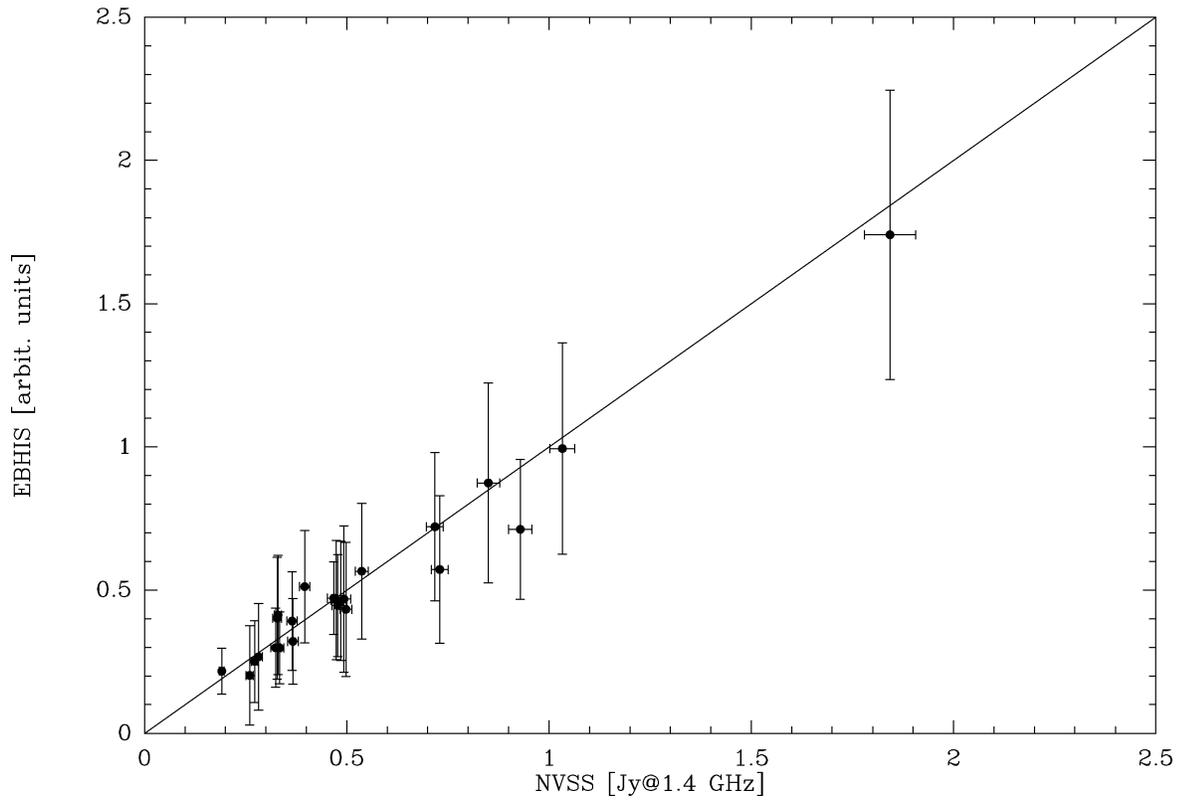}
}
\caption{Comparison of the EBHIS versus the NVSS fluxes of the unresolved radio continuum sources shown in Fig.\ref{contsources}. The radio continuum intensity extracted from the FPGA spectrometer data is consistent with the NVSS fluxes. The error-bars mark the $1-\sigma$ uncertainties. This allows to use EBHIS as a resource for radio continuum emission. Towards the Galactic plane we observed also diffuse radio continuum emission.\label{radcont}}
\end{figure*}
\begin{figure*}
\includegraphics[scale=0.6]{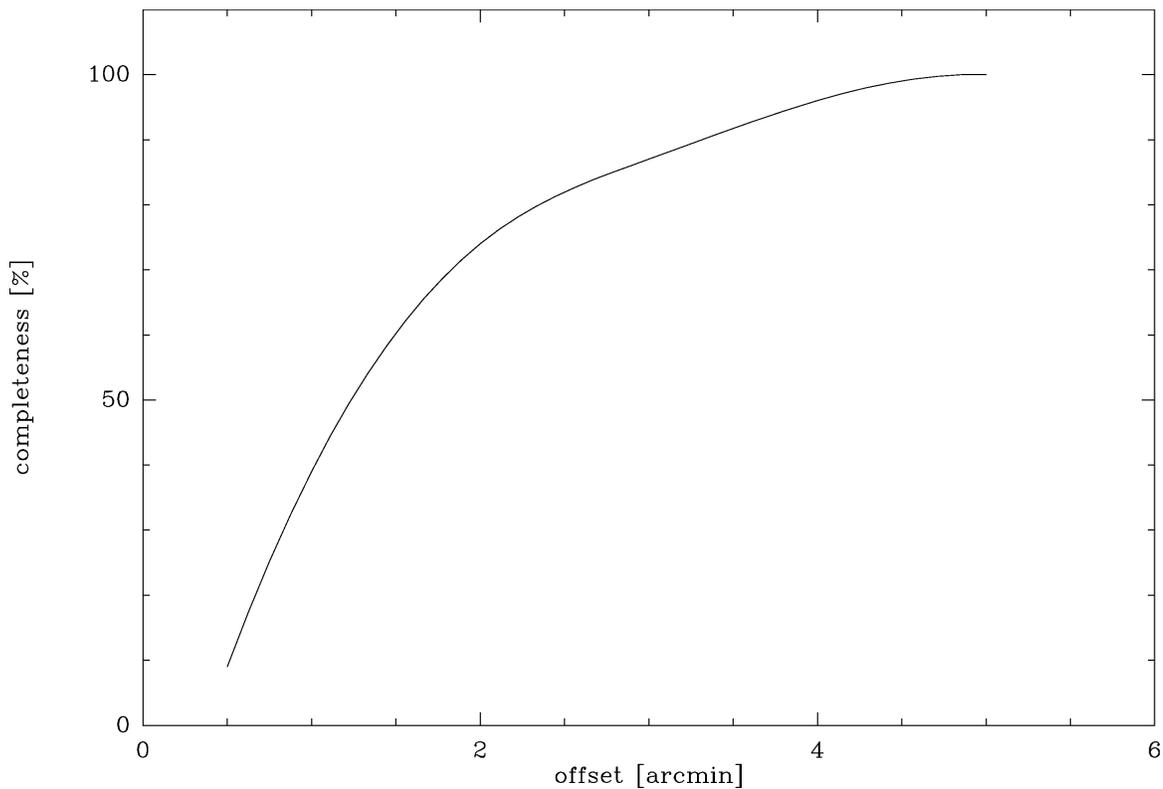}
\caption{Using the radio continuum information it is feasible to monitor the positional accuracy of the final data cube calculation. 74\% of all radio continuum sources are closer to their NVSS position than two arcmin. Two arcmin is also the angular distance of the on--the--fly mapping separation between two consecutive spectral dumps (0.5\,seconds). 100\% of all radio continuum sources are detected within a radius of $5\arcmin$! .\label{contpositions}}
\end{figure*}

\end{document}